\documentclass[pra,preprint,showpacs]{revtex4-1}

\usepackage{amsmath,latexsym,amssymb,amsfonts}
\usepackage[usenames,dvipsnames]{color}
\usepackage[colorlinks]{hyperref}

\usepackage{graphicx}
\usepackage{bm}%for bold Greek letters -- is this used?
\usepackage[applemac]{inputenc}
\usepackage[T1]{fontenc} 

\newcommand{\beq}{\begin{equation}}
\newcommand{\eeq}{\end{equation}}
\newcommand{\bea}{\begin{eqnarray}}
\newcommand{\eea}{\end{eqnarray}}

\newcommand{\eps}{\epsilon}

\begin{document}

\preprint{UdeM-GPP-TH-13-226}  
\preprint{CQUeST-2013-0622}
\preprint{YITP-13-82}

\title{The Battle of the Bulge: Decay of the Thin, False  Cosmic String}
\author{Bum-Hoon Lee$^{a,b}$}
\email{bhl@sogang.ac.kr}
\author{Wonwoo Lee$^{b}$}
\email{warrior@sogang.ac.kr}
\author{Richard MacKenzie$^{c}$}
\email{richard.mackenzie@umontreal.ca}
\author{M.~B.~Paranjape$^{c}$}
\email{paranj@lps.umontreal.ca}
\author{U.~A.~Yajnik$^{d}$}
\email{yajnik@iitb.ac.in}
\author{Dong-han Yeom$^{b,e}$}
\email{innocent.yeom@gmail.com}
\affiliation{$^a$Department of Physics and BK21 Division, Sogang University, Seoul 121-742, Korea}
\affiliation{$^b$Center for Quantum Spacetime, Sogang University, Seoul 121-742, Korea}
\affiliation{$^c$Groupe de Physique des Particules, Département de physique,Universit\'{e} de Montr\'{e}al, C.~P.~6128, Succursale Centre-ville, Montreal, Qu\'{e}bec, Canada, H3C 3J7}
\affiliation{$^d$Department of Physics, Indian Institute of Technology Bombay, Mumbai, India}
\affiliation{$^e$Yukawa Institute for Theoretical Physics, Kyoto University, Kyoto 606-8502, Japan}

\begin{abstract}
We consider the decay of cosmic strings that are trapped in the false vacuum in a theory of scalar electrodynamics in 3+1 dimensions.  This paper is the 3+1 dimensional generalization of the 2+1 dimensional decay of false vortices which we have recently completed \cite{2}.  We restrict our analysis to the case of thin-walled cosmic strings which occur when large magnetic flux trapped inside the string. Thus the string looks like a tube of fixed radius, at which it is classically stable.  The core of the string contains magnetic flux in the true vacuum, while outside the string, separated by a thin wall, is the false vacuum.  The string decays by tunnelling to a configuration which is represented by a bulge, where the region of true vacuum within, is ostensibly enlarged.  The bulge can be described as the meeting, of a kink soliton  anti-soliton  pair, along the length of the string.  It can be described as a bulge appearing in the  initial string, starting from the string of small, classically stable radius, expanding to a fat string of large, classically unstable (to expansion) radius and then returning back to the string of small radius along its length.  This configuration is the bounce point of a corresponding $O(2)$ symmetric instanton, which we can determine numerically.  Once the bulge appears it explodes in real time. The kink soliton anti-soliton pair recede from each other along the length of the string with a velocity that quickly approaches the speed of light, leaving behind a fat tube.  At the same time the radius of the fat tube that is being formed, expands (transversely) as it is no longer classically stable, converting false vacuum to the true vacuum with ever diluting magnetic field within.  The rate of this expansion is determined by the energy difference between the true vacuum and the false vacuum.   Our analysis could be applied to a network, of cosmic strings formed in the very early universe or vortex lines in a superheated superconductor. 
\end{abstract}

\maketitle

\newpage

\section{Introduction \label{sec1}}

We continue our study of the decay of the false vacuum precipitated by the existence of topological defects in that vacuum \cite{1,2}.  Here we consider the case of cosmic strings in a spontaneously broken $U(1)$ gauge theory, a generalized Abelian Higgs model.    The potential for the complex scalar field  has a local minimum at a nonzero value  and the true minimum is at vanishing scalar field.  We assume the energy density splitting between the false vacuum and the true vacuum is very small. The spontaneously broken vacuum is the false vacuum.   

In a scenario where from a high temperature phase, the theory passes through an intermediate phase of spontaneous symmetry breaking, finally arriving at a true vacuum of unbroken symmetry, it is generic that there will be topologically defects.  The phase of the complex scalar field can wrap an integer number of times around a given line in 3 dimensional space.  The line can be infinite or form a closed loop.  Homologous to the given line there must exist a line of zeros of the scalar field. Where the scalar field vanishes corresponds to the true vacuum.  The corresponding minimum energy configuration is called a cosmic string, alternatively a Nielsen-Olesen  string \cite{3} or a vortex string \cite{4}.   In the scenario that we have described, the true vacuum lies at the regions of vanishing scalar field, thus the interior of the cosmic string is in the true vacuum while the exterior is in the false vacuum.  

In our recent work \cite{2}, we considered the decay of vortices in the strictly two spatial dimensional context.  There, the vortex was classically stable at a given  radius $R_0$.    Through quantum tunelling, the vortex could evolve to a larger vortex of radius $R_1$, which was no longer classically stable.  Dynamically the interior of the vortex was at the true vacuum, thus energetically lower by the energy density splitting multiplied by the area of the vortex.  The gain in energy behaved as $\sim R^2$, while the magnetic field energy behaved like $\sim 1/R^2$ and the energy in the wall behaved like $\sim R$.  Thus the energy functional had the form
\beq
E=\alpha/R^2 +\beta R -\epsilon R^2.
\eeq
For sufficiently small $\epsilon$, this energy functional is  dominated by the first two terms.  It is infinitely high for a small radius due to the magnetic energy, and will diminish to a local minimum when the linear wall energy begins to become important.  This will occur at a radius $R_0$, well before the quadratic area energy, due to the energy splitting between the false vacuum and the true vacuum becomes important, for $\eps$ is sufficiently small.  Clearly though, for large enough radius of the thin wall string configuration,  the energy splitting will be the most important term, and a thin walled vortex configuration of sufficiently large radius will be unstable to expanding to infinite radius. However, a vortex of radius $R_0$ will be classically stable and only susceptible to decay via quantum tunnelling. The amplitude for such tunnelling, in the semi-classical approximation, has been calculated in \cite{2}.  

In this paper we consider the generalization of the model to 3+1 dimensions. Here the vortex can be continued along the third, additional dimension as a string, often called a cosmic string.  The interior of the string contains a large magnetic flux distributed over a region of the true vacuum.  It is separated from the region outside by a thin wall, where the scalar field is in the false vacuum.  The analysis of the decay of two dimensional vortices cannot directly apply to the decay of the cosmic string, as the cosmic string must maintain continuity along its length.  Thus the radius of the string at a given position cannot spontaneously make the quantum tunnelling  transition to the larger iso-energetic radius, called $R_1$, without being continuously connected to the rest of the string.  The whole string could in principle spontaneously tunnel to the fat string along its whole length, but the probability of such a transition is strictly zero for an infinite string, and correspondingly small for a closed string loop.  The aim of this paper is to describe the tunnelling transition to a state that corresponds to a spontaneously formed bulge in the putatively unstable thin string.

\section{Energetics and dynamics of the thin, false string \label{sec2}}

\subsection{Set-up}

We consider the abelian Higgs model (spontaneously-broken scalar electrodynamics) with a modified scalar potential corresponding to our previous work \cite{2} but now generalized to 3+1 dimensions.  The Lagrangian density of the model has the form
\beq
{\cal L} = - \frac{1}{4} F_{\mu\nu}F^{\mu\nu} + (D_{\mu}\phi)^*(D^{\mu}\phi)-V(\phi^*\phi),
\label{lagran01}
\eeq
where
$F_{\mu\nu} = \partial_{\mu}A_{\nu} - \partial_{\nu}A_{\mu}$ and
$D_{\mu}\phi = (\partial_{\mu} - ie A_{\mu})\phi$.
The potential is a sixth-order polynomial in $\phi$
\cite{1, pjs}, written
\beq
V(\phi^*\phi) = \lambda(|\phi |^2-\eps v^2) (|\phi |^2-v^2)^2. \label{potential1}
\eeq
Note that the Lagrangian is no longer renormalizable in 3+1 dimensions, however the understanding is that it is an effective theory obtained from a well defined renormalizable fundamental Lagrangian. The fields $\phi$ and $A_\mu$, the vacuum expectation value $v$ have mass dimension 1,  the charge $e$ is dimensionless and $\lambda$ has mass dimension 2 since it is the coupling constant of the sixth order scalar potential. The potential energy density of the false vacuum $|\phi |=v$ vanishes, while that of the true vacuum has $V(0)=-\lambda v^6\eps$.   We rescale analogous to \cite{2} 
\beq
\phi\rightarrow v\phi\quad A_\mu\rightarrow vA_\mu \quad e\rightarrow\lambda^{1/2}ve\quad x\rightarrow x/(v^2\lambda^{1/2})
\eeq
so that all fields, constants and the spacetime coordinates become dimensionless, then the Lagrangian density is still given by Eqn. (\ref{lagran01}) where now the potential is 
\beq
V(\phi^*\phi) = (|\phi |^2-\eps) (|\phi |^2-1)^2. \label{potential}
\eeq
and there is an overall factor of $1/(\lambda v^2)$ in the action.

Initially, the cosmic string will be independent of $z$ the coordinate along its length and will correspond to a tube of radius $R$ with a trapped magnetic flux in the true vacuum inside, separated by a thin wall from the false vacuum outside.  $R$ will vary in Euclidean time $\tau$ and in $z$ to yield an instanton solution.  Thus we promote $R$ to a field $R\rightarrow R(z,\tau)$.  Hence we will look for axially-symmetric solutions for $\phi$ and $A_{\mu}$ in
cylindrical coordinates $(r$, $\theta$, $z$, $\tau)$. We use the following  ansatz
for a vortex of winding number $n$:
\beq
\phi(r, \theta, z, \tau) = f(r, R(z,\tau)) e^{in\theta}, \qquad  A_{i}(r, 
\theta ,z, \tau)=-\frac{n}{e} \frac{\varepsilon^{ij}{r}_j }{r^2}a(r,R( z,\tau)),
\label{ansatz}
\eeq
where $\varepsilon^{ij}$ is the
two-dimensional Levi-Civita symbol.  This ansatz is somewhat simplistic, it is clear that if the radius of the cosmic string swells out at some range of $z$, the magnetic flux will dilute and hence through the (Euclidean) Maxwell's equations some ``electric'' fields will be generated.  In 3 dimensional, source free, Euclidean electrodynamics, there is no distinct electric field, the Maxwell equations simply say that the 3 dimensional magnetic field is divergence  free and rotation free vector field that satisfies superconductor boundary conditions at the location of the wall.  It is clear that the correct form of the electromagnetic fields will not simply be  a diluted magnetic field that always points along the length of the cosmic string as with our ansatz, however the correction will not give a major contribution, and we will neglect it.  Indeed, the induced fields will always be smaller by a power of $1/c^2$ when the usual units are used.  

The Euclidean action functional for the cosmic string then has the form
\bea
S_E[A_{\mu}, \phi]&=& \frac{1}{\lambda v^2}\int d^4x\left[ \sum_i \left(\frac{1}{2}F_{0i}F_{0i}+\frac{1}{2}F_{i3}F_{i3}\right) +\frac{1}{2}F_{03}F_{03}  + \sum_{ij}\frac{1}{4}F_{ij}F_{ij} \right. \nonumber\\
&+& \left.  (\partial_{\tau}\phi)^*(\partial_{\tau}\phi)+(\partial_{z}\phi)^*(\partial_{z}\phi)+\sum_i D_{i}(\phi)^*(D_{i}\phi) + V(\phi^*\phi) \right]  \label{efun}
\eea
where $i,j$ take values just over the two transverse directions and we have already incorporated that $A_0=A_3=0$.  

Substituting Eqns. (\ref{potential},\ref{ansatz}) into Eqn. (\ref{efun}), we obtain
\bea
S_E&=& \frac{2\pi}{\lambda v^2}\int dzd\tau \int^{\infty}_{0} dr\, r \left[ \frac{n^2 \dot{a}^2}{2 e^2r^2}
 + \frac{n^2a'^2}{2 e^2r^2} + \frac{n^2(\partial_r a)^2}{2 e^2r^2} + \dot{f}^2 + f'^2+(\partial_r f)^2\right. \nonumber\\&+&\left. \frac{n^2}{r^2}(1-a)^2f^2
+(f^2-\eps) (f^2-1)^2 \right] ,
\eea
where the dot and  prime denote differentiation with respect to $\tau$ and $z$, respectively.  Then $\dot a =\left(\frac{\partial a(r,R) }{\partial R} \right)  \dot R$ and $a'=\left(\frac{\partial a(r,R) }{\partial R} \right)  R'$, and likewise for $f$, hence the action becomes
\bea\nonumber
S_E&=& \frac{2\pi}{\lambda v^2}\int dzd\tau \int^{\infty}_{0} dr\, r \left[ \frac{n^2 \left(\left(\frac{\partial a(r,R) }{\partial R} \right) \dot R\right)^2}{2 e^2r^2}+ \frac{n^2\left(\left(\frac{\partial a(r,R) }{\partial R} \right)  R'\right)^2}{2 e^2r^2}+ \frac{n^2(\partial_r a)^2}{2 e^2r^2}  \right.\\
 &+&\left. \left(\frac{\partial f(r,R) }{\partial R}  \dot R\right)^2 + \left(\frac{\partial f(r,R) }{\partial R}  R'\right)^2 +(\partial_r f)^2+\frac{n^2}{r^2}(1-a)^2f^2
+(f^2-\eps) (f^2-1)^2 \right] \nonumber\\
&=&  \frac{2\pi}{\lambda v^2}\int dz \int^{\infty}_{0} dr\, r \left[ \left(\frac{n^2 }{2 e^2r^2}\left(\frac{\partial a(r,R) }{\partial R} \right)^2
 + \left(\frac{\partial f(r,R) }{\partial R}  \right)^2\right) (\dot R^2+R'^2)\right.\nonumber\\
 &+&\left. \frac{n^2(\partial_r a)^2}{2 e^2r^2}+(\partial_r f)^2+ \frac{n^2}{r^2}(1-a)^2f^2
+(f^2-\eps) (f^2-1)^2 \right].
\eea
We note the two Euclidean dimensional, rotationally invariant form $(\dot R^2+R'^2)$ which appears in the kinetic term.  This allows to make the $O(2)$ symmetric ansatz for the instanton, and the easy continuation of the solution to Minkowski time, to a relativistically invariant $O(1,1)$ solution, once the tunnelling transition has been completed.

In the thin wall limit, the Euclidean action can be evaluated essentially analytically, up to corrections which are smaller by at least one power of $1/R$.  The method of evaluation is identical to that in \cite{2}, we shall not repeat the details, we find
\beq
S_E=\frac{1}{\lambda v^2}\int d^2x \frac{1}{2}M(R(z,\tau))(\dot R^2+R'^2) +E(R(z,\tau))-E(R_0)
\eeq
where
\bea
M(R)&=&\left[ \frac{2\pi n^2}{e^2R^2}+\pi R  \right] \\
E(R)&=&\frac{n^2\Phi^2}{2\pi R^2}+\pi R-\eps\pi R^2
\eea
and $R_0$ is the classically stable thin tube string radius. 
\begin{figure}[ht]
\centerline{\includegraphics[width=0.9\linewidth]{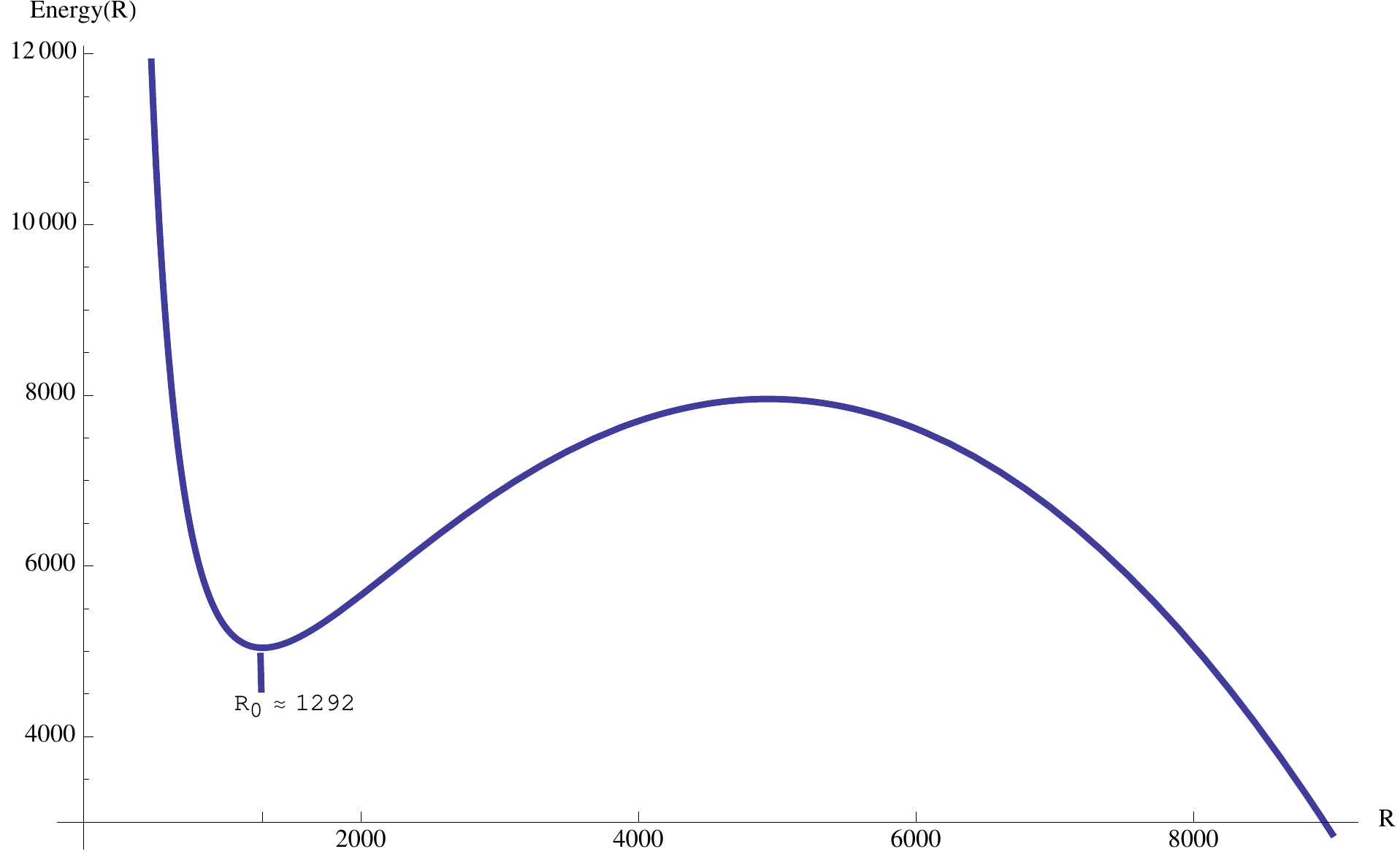}} \caption{\label{energy} (color online) The energy as a function of $R$, for $n=100, e=.005$ and $\epsilon=.001$.}
\end{figure}
\section{Instantons and the bulge}
\subsection{Tunnelling instanton}
We look for an instanton solution that is $O(2)$ symmetric, the appropriate ansatz is
\beq
R(z,\tau)=R(\sqrt{z^2+\tau^2})=R(\rho)
\eeq
with the imposed boundary condition that $R(\infty)=R_0$. 

\begin{figure}[ht]
\centerline{\includegraphics[width=0.9\linewidth]{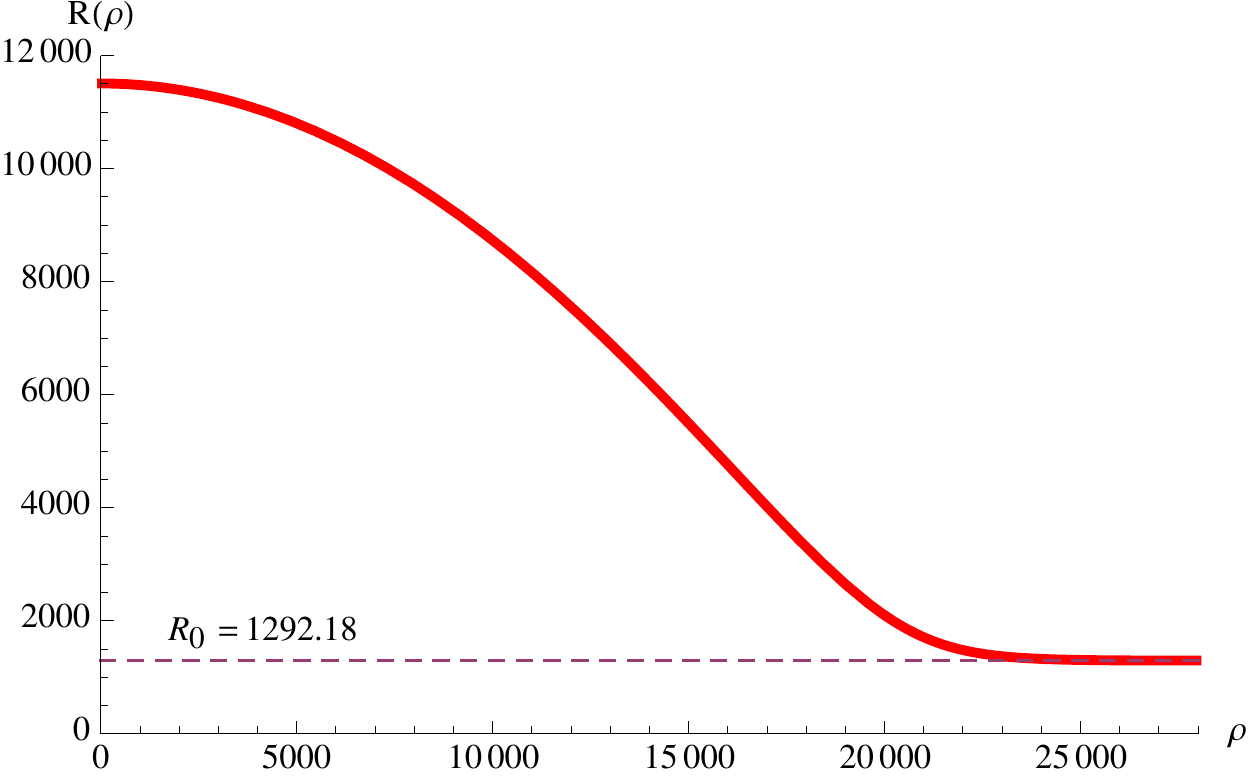}} \caption{ (color online) The radius as a function of $\rho$.}\label{radius}
\end{figure}

Such a solution will describe the transition from a string of radius $R_0$ at $\tau=-\infty$, to a point in $\tau=\rho_0$ say at $z=0$ when a soliton anti-soliton pair is started to be created.  The configuration then develops a bulge which forms when the pair separates to a radius which has to be again $\rho_0$ because of $O(2)$ invariance and which is the bounce point of the instanton  along the $z$ axis at $\tau=0$.  Finally the  subsequent Euclidean time evolution continues in a manner which is just the  (Euclidean) time reversal of evolution leading up to the bounce point configuration  until a simple cosmic string of radius $R_0$ is re-established for $\tau\ge\rho_0$ and all $z$, {\it i.e.} $\rho\ge\rho_0$.  The action functional is given by
\beq
S_E=\frac{2\pi}{\lambda v^2}\int d\rho\,\, \rho \left[ \frac{1}{2}M(R(\rho))\left(\frac{\partial R(\rho)}{\partial\rho}\right)^2 +E(R(\rho))-E(R_0)\right] .
\eeq
The instanton equation of motion is
\beq
\frac{d}{d\rho}\left(\rho M(R)\frac{dR}{d\rho}\right) -\frac{1}{2}\rho M'(R)\left(\frac{dR}{d\rho}\right)^2-\rho E'(R)=0
\eeq
with the boundary condition that $R(\infty)=R_0$, and we look for a solution that has $R\approx R_1$ near $\rho =0$.  The solution necessarily ``bounces'' at $\tau=0$ since $\partial R(\rho)/\partial\tau|_{\tau=0}=R'(\rho) (\tau/\rho)|_{\tau=0}=0$. (The potential singularity at $\rho=0$ is not there since a smooth configuration requires $R'(\rho)|_{\rho =0}=0$.)  The equation of motion is better cast as an essentially conservative dynamical system with a ``time'' dependent mass and the potential given by the inversion of the energy function as pictured in Fig. \eqref{energy}, but in the presence of a ``time'' dependent friction where $\rho$ plays the role of time:
\beq
\frac{d}{d\rho}\left(M(R)\frac{dR}{d\rho}\right) - \frac{1}{2}M'(R)\left(\frac{dR}{d\rho}\right)^2- E'(R)=-\frac{1}{\rho}\left(M(R)\frac{dR}{d\rho}\right).\label{eqofm}
\eeq
As the equation is ``time'' dependent, there is no analytic trick to evaluating the bounce configuration and the corresponding action.  We are, however,  confident in the existence of a solution which starts with a given $R\approx R_1$ at $\rho=0$ and achieves $R=R_0$ for $\rho>\rho_0$, by showing the existence of an initial condition that gives an overshoot and another initial condition that gives an undershoot, in the same manner of proof as in \cite{Coleman}.  If we start at the origin at $\rho=0$ high enough on the right side of the inverted energy functional  pictured in Fig. \eqref{energy}, the equation of motion \eqref{eqofm}  will cause the radius $R$, to slide down the potential and then roll up the hill to $R=R_0$.  If we start too far up to the right, we will roll over the maximum at $R=R_0$ while if we do not start high enough we will never make it to the top of the hill at $R=R_0$.  The RHS of \eqref{eqofm} acts as a ``time'' dependent friction, which becomes negligible as $\rho\to\infty$, and once it is negligible, the motion is effectively conservative.  We resort to numerical studies and we find with little difficulty, that if we start at $R\approx 11506.4096$, for $n=100, e=.005$ and $\epsilon=.001$ we generate the profile function $R(\rho )$ in Figure \eqref{radius}.
Actually, numerically integrating to $\rho\approx 80,000$ the function falls back to the minimum of the inverted energy functional Eqn. \eqref{energy}.  On the other hand, we increase the starting point by $.0001$, the numerical solution overshoots the maximum at $R=R_0$.  Hence we have numerically implemented the overshoot/undershoot criterion of \cite{Coleman}.  

\begin{figure}
\begin{center}
\includegraphics[width=7cm]{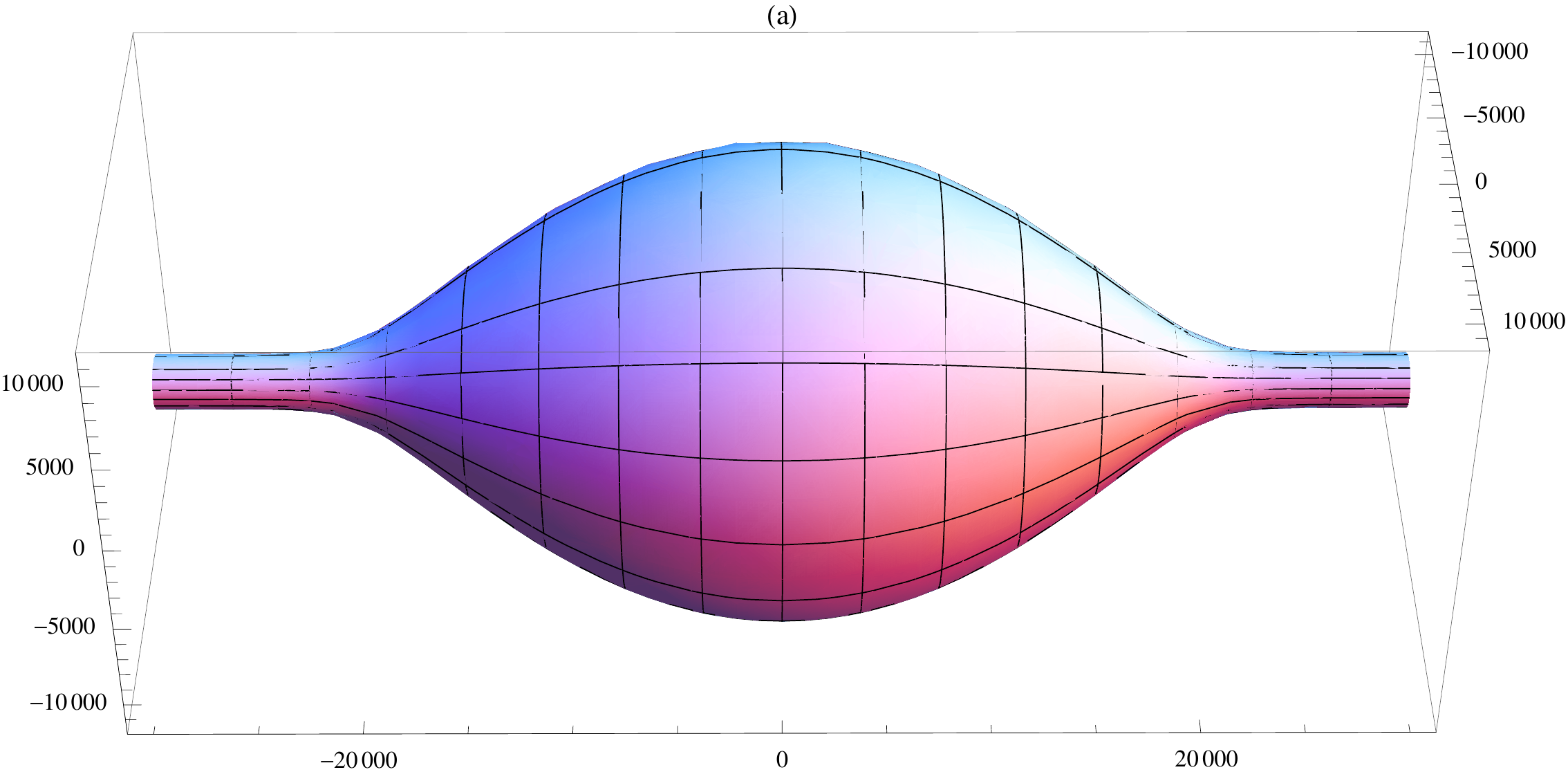}
\qquad
\includegraphics[width=8cm]{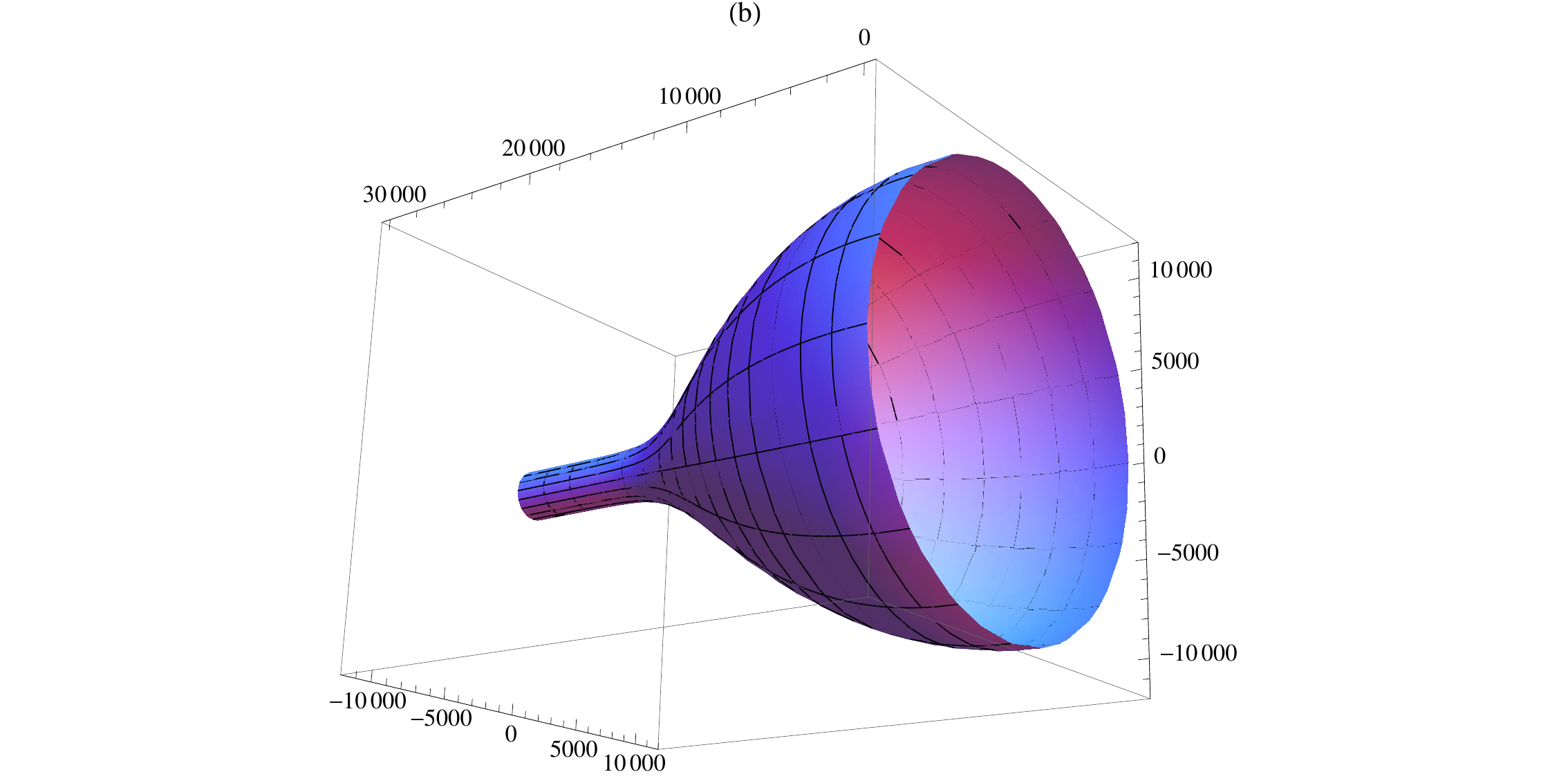}
\end{center}
\caption{(Color online) (a) Cosmic string  profile at  the bounce point.  (b) Cut away of the cosmic string profile at bounce point. } 
\label{bp}
\end{figure}

The cosmic string emerges with a bulge described by the function numerically evaluated and represented in Figure \eqref{radius} which corresponds to $R(z, \tau=0)$.  A 3 dimensional depiction of the bounce point is given in Figure \eqref{bp}.  One should imagine the radius $R(z)$ along the cosmic string  to be  $R_0$ to the left, then bulging out to the the large radius as described by the mirror image of the function in Figure \eqref{radius} and then returning to $R_0$ according to the function in Figure \eqref{radius}.  This radius function has argument $\rho=\sqrt{z^2+\tau^2}$.  Due to the Lorentz invariance of the original action, the subsequent Minkowski time evolution is given by $R(\rho)\to R(\sqrt{z^2-t^2})$, which is of course only valid for $z^2-t^2\ge 0$. Fixed $\rho^2=z^2-t^2$ describes a space-like hyperbola that asymptotes to the light cone.  The value of the function $R(\rho)$ therefore remains constant along this hyperbola.  This means that the point at which the string has attained the large radius moves away from $z\approx 0$ to $z\to\infty$ at essentially the speed of light.  The other side of course moves towards $z\to -\infty$.  Thus the soliton anti-soliton pair separates quickly moving at essentially the speed of light, leaving behind a fat cosmic string, which is subsequently, classically unstable to expand and fill all space.  

\subsection{Tunnelling amplitude}
It is difficult to say too much  about the tunnelling amplitude or the decay rate per unit volume analytically in the parameters of the model.  The numerical solution we have obtained for some rather uninspired choices of the parameters gives rise to the profile of the instanton given in Fig. \eqref{radius}.  This numerical solution could be then inserted into the Euclidean action to determine its numerical  value, call it $S_0(\eps)$.  It seems difficult to extract any analytical dependence on $\epsilon$, however it is reasonable to expect that as $\eps\to 0$ the tunnelling barrier, as can be seen in Fig. \eqref{energy}, will get bigger and bigger and hence the tunnelling amplitude will vanish. On the other hand, there should exist a limiting value, call it $\eps_c$, where  the tunnelling barrier disappears at the so-called dissociation point \cite{u}, such that as  $\eps\to\eps_c$,  the action of the instanton will vanish, analogous to what was found in \cite{2}.  In general the decay rate per unit length of the cosmic string will be of the form
\beq
\Gamma= A^{\rm c.s.} \left(\frac{S_0(\eps)}{2\pi}\right)e^{-S_0(\eps)}.
\eeq
where $A^{\rm c.s.}$ is the determinantal factor excluding the zero modes and $\left(\frac{S_0(\eps)}{2\pi}\right)$ is the correction obtained after taking into account the two zero modes of the bulge instanton.  These  correspond to invariance under Euclidean time translation  and   spatial translation along the cosmic string \cite{Coleman}. 
In general, there will be  a length $L$ of cosmic string per volume $L^3$.  For a second order phase transition to the metastable vacuum, $L$ is the correlation length at the temperature of the transition which satisfies $L^{-1}\approx \lambda v^2 T_c$ \cite{kz}.  For first order transitions, it is not clear what the density of cosmic strings will be.  We will keep $L$ as a parameter but we do expect that it is microscopic.  Then in a large volume $\Omega$, we will have a total length $N L$ of cosmic string, where $N=\Omega/L^3$.  Thus the decay rate for the volume $\Omega$ will be
\beq
\Gamma \times (NL)=\Gamma \left(\frac{\Omega}{L^2}\right)= A^{\rm c.s.} \left(\frac{S_0(\eps)}{2\pi}\right)e^{-S_0(\eps)}\frac{\Omega}{L^2}
\eeq
or the decay rate per unit volume will be
\beq
\frac{\Gamma}{L^2}=\frac{A^{\rm c.s.} \left(\frac{S_0(\eps)}{2\pi}\right)e^{-S_0(\eps)}}{L^2}.
\eeq
A comparable calculation with point-like defects \cite{2} would give a decay rate per unit volume of the form
\beq
\frac{\Gamma^{\rm point\, like} }{L^3}=\frac{A^{\rm point\, like} \left(\frac{S_0^{\rm point\, like}(\eps)}{2\pi}\right)^{3/2}e^{-S_0^{\rm point\, like}(\eps)}}{L^3}
\eeq
and the corresponding decay rate from vacuum bubbles (without topological defects) \cite{Coleman} would be
\beq
\Gamma^{\rm vac.\, bubble} =A^{\rm vac.\, bubble} \left(\frac{S_0^{\rm vac.\, bubble} (\eps)}{2\pi}\right)^2e^{-S_0^{\rm vac.\, bubble} (\eps)}.
\eeq
Since the length scale $L$ is expected to be microscopic,  we would then find that the number of defects in a macroscopic volume ({\it i.e.} universe) could be incredibly large, suggesting that the decay rate from topological defects would dominate over the decay rate obtained from simple vacuum bubbles à la Coleman \cite{Coleman}.   Of course the details do depend on the actual values of the Euclidean action and the determinantal factor that is obtained in each case.   
\section{Conclusion}
There are many instances where the vacuum can be meta-stable.  The symmetry broken vacuum can  be  metastable.  Such solutions for the vacuum can be important for cosmology and for the case of supersymmetry breaking see \cite{abel} and the many references therein.  In string cosmology, the inflationary scenario that has been obtained in\cite{kklt},   also gives rise to a vacuum that is  meta-stable , and it must necessarily be long-lived to have cosmological relevance.  

In a condensed matter context symmetry breaking ground states are also of great importance.  For example, there are two types of superconductors \cite{am}. The cosmic string is called a vortex line solution in this context, and it is relevant to type II superconductors.   The vortex line contains an unbroken symmetry region that carries a net magnetic flux, surrounded by a region of broken symmetry.  If the temperature is raised, the true vacuum becomes the unbroken vacuum, and it is possible that the system exists in a superheated state where the false vacuum is meta-stable \cite{dolgert}.  This technique has actually been used to construct detectors for particle physics \cite{pd}.  Our analysis might even describe the decay of vortex lines in superfluid liquid 3Helium \cite{legg}.

The decay of all of these metastable states could be described through the tunnelling transition mediated by instantons in the manner that we have computed in this article.  For appropriate limiting values of the parameters, for example when $\eps\to\eps_c$, the suppression of tunnelling is absent, and the existence of vortex lines or cosmic strings could cause the decay of the meta-stable vacuum without bound.  Experimental observation of this situation would be interesting.

\section*{Acknowledgements}
This work was financially supported in part by the National Research Foundation of Korea grant funded by the Ministry of Education, Science and Technology through the Center for Quantum Spacetime (CQUeST) of Sogang University (2005-0049409), by the Natural Science and Engineering Council of Canada, by the Department of Science and Technology, India, by  the Coopération Québec-Maharashtra (Inde) program of the Ministère des relations internationales du Québec and by the Direction de relations internationales de l'Université de Montréal. WL was supported by the Basic Science Research Program through the National Research Foundation of Korea (NRF) funded by the Ministry of Education, Science and Technology (2012R1A1A2043908). DY is supported by the JSPS Grant-in-Aid for Scientiﬁc Research (A) No. 21244033. RM thanks McGill University for hospitality while this work was in progress.

\end{document}